\documentstyle[aps,prd,twocolumn,epsf]{revtex}
\tighten
\newcommand \beq{\begin{eqnarray}}
\newcommand \eeq{\end{eqnarray}}
\newcommand \ga{\raisebox{-.5ex}{$\stackrel{>}{\sim}$}}
\newcommand \la{\raisebox{-.5ex}{$\stackrel{<}{\sim}$}}

\begin{document}
\title{Color Superconducting Quark Matter in Neutron Stars}
\author{Henning Heiselberg}
\address{NORDITA, Blegdamsvej 17, DK-2100 Copenhagen \O, Denmark}

\maketitle

\begin{abstract}
Color superconductivity in quark matter is studied for electrically
charge neutral neutron star matter in $\beta$-equilibrium. Both bulk
quark matter and mixed phases of quark and nuclear matter are
treated. The electron chemical potential and strange quark mass affect
the various quark chemical potentials and therefore also the
color superconductivity due to dicolor pairing or color-flavor
locking.
\end{abstract}

\vspace{2cm}

Strongly interacting matter is expected to undergo a transition to
chirally restored matter of quarks and gluons at sufficiently high
baryon or energy density.  Such phase transitions are currently
investigated in relativistic heavy-ion collisions and may exist in the
interior of neutron stars.  At low temperatures a condensate of quark
Cooper pairs may appear characterized by a BCS gap $\Delta$
\cite{Love,Wilczek,Shuryak,ABR,Carter} usually referred to as color
superconductivity (CSC).  The appearance of a gap through color-flavor
locking (CFL) requires the gap to exceed the difference between the
quark Fermi momenta, which is not the case for sufficiently
large strange quark masses. In neutron star matter the presence of an
appreciable electron chemical potential, $\mu_e$, also change the
conditions for CFL as discussed in the following.

In neutron star matter $\beta$-equilibrium relates the quark
and electron chemical potentials
\beq
   \mu_d=\mu_s=\mu_u+\mu_e \,. \label{chem}
\eeq
Temperatures are normally much smaller than typical Fermi energies in
neutron stars. If
interactions are weak, 
the chemical potentials are then related to Fermi momenta
by $\mu_i=\sqrt{m_i^2+p_i^2}$. 
If the strange quark mass $m_s$ is  much smaller than the 
quark chemical potentials, Eq. (\ref{chem}) implies
a difference between
the quark Fermi momenta
\begin{eqnarray}
   p_u-p_d &=& \mu_e                           \,, \label{ud} \\
   p_u-p_s &\simeq& \frac{m_s^2}{2\mu} -\mu_e  \,, \label{us} \\
   p_d-p_s &\simeq& \frac{m_s^2}{2\mu}         \,, \label{ds} 
\end{eqnarray}
where $\mu$ is an average quark chemical potential.
In Fig. 1 the difference between the Fermi momenta given by
Eqs. (\ref{ud}-\ref{ds}) are plotted as function of electron chemical
potential. 
Strange quark masses are estimated from low energy QCD
$m_s\simeq150-200$~MeV and typical quark chemical potentials are
$\mu\simeq 400-600$~MeV in quark matter \cite{physrep}. Consequently,
$m_s^2/2\mu\simeq 10-25$~MeV.

\begin{figure}
\epsfxsize=8.6truecm 
\epsfbox{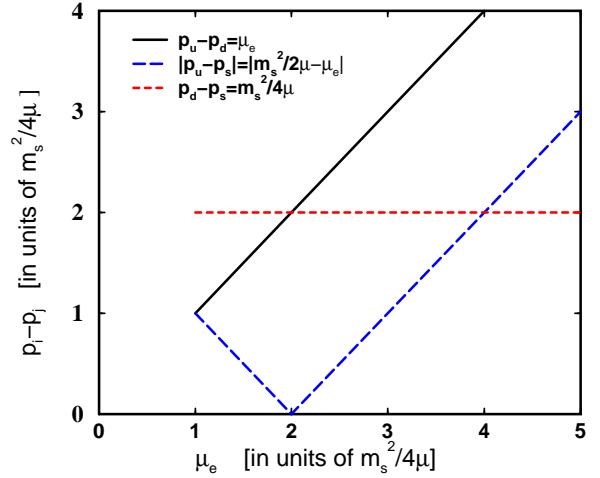}
\caption[]{The difference between Fermi momenta as function of
electron chemical potentials. All quantities in units of 
$m_s^2/4\mu$. At densities where the difference exceeds the CSC gap the
color-flavor locking is broken and CSC disappears.}
\label{cscfig}
\end{figure}

Perturbative corrections change the relation between Fermi momenta and
chemical potentials for relativistic
quarks to $p_q=\mu_q(1-2\alpha_s/3\pi)^{1/3}$ and lead to 
corrections of order $\alpha_sm_s^2/\mu$ 
in Eqs. (\ref{us}-\ref{ds}) for a massive strange quark. 
For weak coupling and small strange quark
masses such effects are small and will be ignored in the following.

The BCS gap equation has previously been solved for {\em u,d} and
{\em u,d,s} quark matter ignoring electrons and $\beta$-equilibrium 
and the conditions for condensates of dicolor pairs and CFL
respectively were obtained \cite{Love,Wilczek,Shuryak,ABR}.
The CFL condition consists of three pair-wise ``CFL'' conditions
\beq 
  \Delta > |p_i-p_j| \,,\quad i,j=u,d,s \label{CFL} \,, 
\eeq
and thus require both small electron chemical potential and strange
quark mass according to Eqs. (\ref{ud}-\ref{ds}).  Alternatively, only
one of these conditions may be fulfilled.  For example, {\it ud}
``CFL'' requires small electron chemical potential whereas {\it ds}
``CFL'' requires a small strange quark mass.  The {\it us} ``CFL''
condition can actually be satisfied when $\mu_e\simeq m_s^2/2\mu$.
For these three cases a condensate of dicolor pairs (2CS) can appear
between {\it ud, us, ds}-quarks analogous to the standard {\it ud} 2CS
usually discussed for symmetric {\it ud} quark matter
\cite{Wilczek,Shuryak}.

The magnitude of the electron chemical potential will now be discussed for
electrically charge neutral bulk quark matter as well as for 
mixed phases of quark and nuclear matter \cite{HPS}.

{\bf Bulk quark matter} must be electrically neutral, i.e., the
net positively charged quark density must be balanced by the
electron density
\beq
  n_e&=&\frac{\mu_e^3}{3\pi^2}=\frac{2}{3}n_u-\frac{1}{3}n_d-\frac{1}{3}n_s 
         \nonumber\\
  &\simeq& \frac{1}{\pi^2}\left(\frac{1}{2}m_s^2\mu-2\mu_e\mu^2\right)
                                 \,. \label{bulk}
\eeq
Muons will appear when their chemical potential exceeds their
rest masses, $\mu_\mu=\mu_e>m_\mu$, but this occurs for very
large electron chemical potentials only, where CSC is unlikely,
and muons will therefore be ignored here.

When the electron chemical potential is small as compared to the quark
chemical potentials the l.h.s. of Eq. (\ref{bulk}) is negligible and
we obtain
\beq
   \mu_e \simeq \frac{m_s^2}{4\mu} \,. \label{mue}
\eeq
Albeit the electron charge density is negligible, the electrons
affect quark chemical potentials through $\beta$-equilibrium. 
The ``CFL'' condition for {\it ud}-quarks, Eq. (\ref{ud}),
is therefore the {\it same} as the ``CFL'' condition for {\it us}-quarks,
Eq. (\ref{us}), in pure quark matter.

{\bf A mixed phase} of quark and nuclear matter has lower energy per
baryon at a wide range of densities \cite{Glendenning}
if the Coulomb and surface energies
associated with the structures are sufficiently small \cite{HPS,physrep}.
The mixed phase will then consist of two coexisting phases of nuclear
and quark matter in droplet, rod- or plate-like structures in
a continuos background of electrons much like the mixed phase of
nuclear matter and a neutron gas in the inner crust of neutron
stars \cite{Lorenz}. Another requirement for a mixed phase is that the
length scales of such structures must be shorter than typical screening
lengths.

In the mixed phase the nuclear and quark matter will be positively and
negatively charged respectively.
Total charge neutrality requires
\beq
  n_e=(1-f)n_p+f(\frac{2}{3}n_u-\frac{1}{3}n_d-\frac{1}{3}n_s) \,, \label{mix}
\eeq
where $n_p$ is the proton density and $f$ is the ``filling fraction'',
i.e. the fraction of the volume filled by quark matter. For pure
nuclear matter, $f=0$, the nuclear symmetry energy can force the
electron chemical potential above $\sim100$ MeV at a few times normal
nuclear matter densities. With increasing filling fraction, however,
negative charged droplets of quark matter replace some of the
electrons and $\mu_e$ decreases. With increasing density and filling
fraction it drops to its minimum value given by Eq.  (\ref{mue})
corresponding to pure quark matter, $f=1$. 

{\bf Gap sizes} of order a few MeV or less were originally estimated
within perturbative QCD \cite{Love}. Non-perturbative calculations
give large gaps of order a few tens of MeV
\cite{Wilczek,Shuryak,ABR,Carter}.  At high densities $\mu\to\infty$ and
weak couplings ($g$) the gap has been calculated, $\Delta\simeq
g^{-5}\exp(-3\pi^2/\sqrt{2}g)$, and drops below $\sim1$~MeV
\cite{Son}. Information about the non-perturbative low density limit
may obtain from studies of dilute Fermi systems. At
low densities, when the typical range of interaction, $R$, is much
shorter than the scattering length, $|a|$, the gap is $\Delta\sim
\mu\exp(-\pi/2p_F|a|)$ \cite{Gorkov}, where $a$ is the
non-relativistic scattering length. 
For $p_F|a|\sim 1$ the gap may be large - of order the Fermi energy.
Large scattering
lengths appear when the two interacting particles almost form a bound
state. However, confinement of quarks is different from such a simple
bound state analogy and the large gap of order the chemical potential
may not be conjectured for relativistic quarks.

In the mixed phase gaps may also be affected by the finite size of the
quark matter structures. For example, nuclei pairing is dominated by
surface effects \cite{BM} since gaps in nuclear matter are larger at
lower densities. As droplets of quark matter are of similar size and
baryon number we may expect similar finite size effects to enhance the
CSC gap sizes.

For large gaps it may also be energetically favorable to have
spatially varying quark chemical potentials and densities such that
CFL occurs in some regions but not in others. From the gain in energy
of order $\Delta^2/\mu$ per particle the system must, however, pay
Coulomb and surface energies associated with these structures
\cite{HPS}.  A similar scenario is considered in \cite{Bedaque}
for {\it u,d} quark matter.

{\bf Consequences}: Some bulk or mixed phase regions of quark matter
in neutron stars can be color superconducting either by CFL or 2CS
depending on the gap sizes, electron chemical potentials and strange
quark masses as described above.  Furthermore, temperatures in neutron
stars are so low, $T\la 10^6K\simeq 10^{-4}$~MeV, that quark matter
structures would be solid frozen. As a consequence, lattice vibration
will couple electrons at the Fermi surface with opposite momenta and
spins via phonons and lead to a ``standard'' BCS gap for electrons.
The isotopic masses are similar but as densities and Debye frequencies
are larger, we can expect considerably larger BCS gaps for
electrons.  At typical neutron star densities neutrons and protons are
superfluid as well due to $^1S_0$ and, in the case of protons, also
$^3P_2$ pairing \cite{physrep}.  These superfluid and superconducting
components will have drastically different transport properties than
normal Fermi liquids \cite{Pethick}.  Generally the resistance,
specific heat, viscosities, cooling, etc. are suppressed by factors of
order $\sim\exp(-\Delta_i/T)$, where $\Delta_i$ is the gap of quarks,
nucleons or electrons.

In relativistic nuclear collisions the strange quark chemical
potential is zero initially and expansion times $\sim R/c\simeq
10$~fm/c are short as compared to time scales for weak decay and
strangeness distillation. Therefore, $\mu_s\simeq 0$ and we expect no
CFL.  In heavy ion collisions the amount on neutrons and therefore
also {\it d}-quarks exceed that of protons and {\it u}-quarks.  The
resulting difference, $|p_d-p_u|$, can prohibit a 2CS depending on
density, temperature and gap size.

{\bf In summary} the conditions for color superconductivity in quark
matter were given in Eqs. )\ref{ud}-\ref{ds}) for electrically charge
neutral neutron star matter in $\beta$-equilibrium - both bulk quark
matter and mixed phases of quark and nuclear matter.  The electron
chemical potential and strange quark mass affect the various quark
chemical potentials. For CFL to occur the gap must exceed both the
electron chemical potential, $\Delta\ga\mu_e$, and the mitch-match in
Fermi momenta induced by a massive strange quark, $\Delta\ga
m_s^2/2\mu$.  Alternatively, if $\mu_e$, $m_s^2/2\mu$ or the
difference $|\mu_e-m_s^2/2\mu|$ are smaller than the gap, then a
condensate of dicolor pairs (2CS) can appear between {\it ud, ds,
us}-quarks respectively analogous to the standard {\it ud} 2CS usually
discussed for symmetric {\it ud} quark matter.

\end{document}